\documentclass[12pt]{article}
\usepackage[cp866]{inputenc}
\begin{document}
\newcommand{\beq}{\begin{equation}}
\newcommand{\eeq}{\end{equation}}
\newcommand{\var}{\varepsilon}
\newcommand{\pr}{\partial}
\begin{center}
{\bf Unconventional  Superconductivity in Two-Dimensional Electron
Systems with Longe-Range Correlations}
\vskip 0.5 cm
  {\it  V.~A. Khodel}

\small{Kurchatov Institute, Russian Research Center, 123182 Moscow,
Russia}

{\it V.~ M. Yakovenko}

{\small University of Maryland, College Park, MD 20472-4111,  USA}
\begin{abstract}
Properties of superfluid states of two-dimensional electron
systems with critical antiferromagnetic fluctuations are
investigated.
These correlations are found  to  result in the emergence of
rapidly varying in the
momentum space  terms in all components of the mass operator, including
 the gap function $\Delta({\bf p})$. It is shown that a domain, where
 these terms reside, shrinks with the temperature, leading to
 a significant difference between the temperature
$T_c$, at which superconductivity is terminated, and the
temperature $T^*$, where the gap in the single-particle spectrum
vanishes.
\end{abstract}
\end{center}

The problem of high-temperature  superconductivity is known  to
defy solution within the  Fermi liquid approach. Initially
it  looked like  it had to do with  only normal states of
high-$T_c$ superconductors. However, later it was acknowledged that
 the BCS theory fails in dealing with  their superfluid states, as well
\cite{shen,camp}. This is best demonstrated by the persistence of
a gap $\Delta$ in the spectra of single-particle (sp) excitations of
many high-$T_c$ superconductors  above
  the critical  temperature $T_c$, at which
superconductivity disappears (the so-called pseudogap phenomenon
\cite{shen,camp,ue1,ue2}).

          Another salient feature of two-dimensional electron liquid
of high-$T_c$ superconductors is the universality of its phase diagram versus the doping $x$. At low
 $|x|\leq x_c\simeq 0.05$, corresponding to the filling, close to $1/2$,
two-dimensional compounds are antiferromagnetic insulators. At larger $x$,
aniferromagnetic ordering is nil, but
 in the vicinity of the  phase transition,
long-range correlations
 with wave vectors ${\bf q}$, close to
 the antiferromagnetic vector  ${\bf Q}=(\pi,\pi)$,  turn out to be
 drastically enhanced, which results in  the divergence  of the
      electron-electron scattering amplitude
$\Gamma=\Gamma_0+\Gamma_a
 {\mathbf \sigma}_1 {\mathbf\sigma}_2$  with
    \beq \Gamma_a({\bf q}\to {\bf Q},\omega\to 0;x)\sim
[({\bf q}-{\bf Q})^2+r_a^{-2}(x)+ic|\omega|]^{-1} \        ,
\label{longa} \eeq
 the correlation radius  $r_a(x)$
becoming infinite at $x=x_c$    \cite{schrief,pines}.

The impact  of this singularity on   sp properties  is studied
  proceeding from the RPA formula  $\Sigma_a=(\Gamma_a*G)$, which
presents an associated
 with antiferromagnetic fluctuations part $\Sigma_a$ of the mass operator
$\Sigma$ as a convolution
of the amplitude $\Gamma_a$ with
 the sp Green function $G$
( see e.g. [5-9]).
For a long time,
attention  was focused on
   the  energy dependence of $\Sigma$,
while its momentum dependent part $\Sigma({\bf p},\var=0)$
was parameterized by the
effective mass $m^*$. This is justified  in systems with  short-range
correlations, where the mass operator  $\Sigma({\bf p},\var=0)$ is a smooth function
of ${\bf p}$. But this is not the case.
 Straightforward calculations
show that long-range correlations (\ref{longa}) trigger a rapidly
varying with ${\bf p}$ component $ \Sigma_a({\bf p})$
of the function $\Sigma({\bf p},\var=0)\equiv \Sigma_r({\bf p})+
\Sigma_a({\bf p})$, being a
 convolution of   $\Gamma_a$ and the pole part $G^q$ of
the Green function  $G$. It
 should be emphasized that  $\Sigma_a({\bf p})$ has to be
evaluated self-consistently, otherwise  the
 flattening of the sp spectra $\xi({\bf p})$ in  normal states,
found in Ref. \cite{ks}
and observed in many high-$T_c$ compounds, gets lost.

To get rid of the energy-dependent terms in $\Sigma$
 we calculate
 the derivative $\partial{\rm Re}\Sigma({\bf p},\var)/ \partial{\bf p}\to
 ({\rm Re}\Gamma_a*
\partial {\rm Im}G^q/\partial{\bf p})$. After simple algebra
we obtain
  \beq {\pr {\rm Re}
\Sigma_a ({\bf p})\over \pr {\bf p}}= {3\over
2}z \int \Gamma_a({\bf p}-{\bf p}_1,\omega=E({\bf p})) {\pr n({\bf p}_1,T)\over
\pr {\bf p}_1}d\tau_1 \ .\label{psig}  \eeq
  Here
$z=[1- \left({\pr \Sigma(\var)\over \pr\var}\right)_0]^{-1}$ is
the renormalization factor, $d\tau=d^2p/(2\pi)^2$,  and
 \beq n({\bf p},T)=v^2({\bf
p})(1-f(E))+ (1-v^2({\bf p}))f(E)={1\over 2}- {\xi({\bf p})\over 2E({\bf p})}
\tanh {E({\bf p})\over 2T}
\label{np} \eeq
is the quasiparticle momentum distribution. In this formula,
 $f(E)=(1+\exp(E/T))^{-1}$, while
$ v^2({\bf p})=\left(E({\bf p}) -\xi({\bf p})\right)/ 2E({\bf
p})$, where $E({\bf p})=\sqrt{\xi^2({\bf p})+\Delta^2({\bf p})}$
and $\Delta$ is the gap function, while $\xi({\bf p})=z\left(\xi^0_{{\bf p}}+\Sigma_a({\bf
p})+\Sigma_r({\bf p})\right)\equiv \xi^0({\bf p})+z\Sigma_a({\bf
p}) $ is the sp energy spectrum of the normal
 state  measured from the chemical potential $\mu$.  To a good approximation,
the spectrum  $\xi^0({\bf p})$ and the LDA electron spectrum
 $\xi^0_{{\bf p}}$ are related by
 $\xi^0({\bf p})= \xi^0_{{\bf p}}/m^*$.

In what follows, the argument  $\omega=E({\bf p}_1)$
   of the function
$\Gamma_a({\bf p}-{\bf p}_1,\omega)$ in Eq. (\ref{psig})  is replaced by 0,
since both the functions $\pr n({\bf p}_1,T)/ \pr {\bf p}_1$  and
$\Gamma_a({\bf p}-{\bf p}_1)$, taken at $\xi({\bf p})=0$, are
peaked at $ \xi({\bf p}_1)=0$. Upon inserting this result into
Eq.(\ref{psig}) and integrating over momenta, one finds
 \beq \xi
({\bf p})= \xi^0({\bf p})+ {3\over 2}z^2 \int \Gamma_a({\bf p}-{\bf
p}_1)
 n({\bf p}_1,T)d\tau_1 \  .
\label{sig} \eeq

The  gap $\Delta({\bf p})$, obeying   the BCS gap equation, is also   decomposed
 into a sum $\Delta({\bf p})=\Delta_a({\bf p})+\Delta_r({\bf p})$ of a
regular $\Delta_r({\bf p})$
 and a rapidly varying with ${\bf p}$ component $\Delta_a({\bf p})$. In the
 case of singlet pairing,
 the respective  equation for $\Delta_a({\bf p})$ reads \cite{mig}:
\beq \Delta_a({\bf p})= -3z^2 \int \Gamma_a({\bf p}-{\bf
p}_1) {\tanh{E({\bf
p}_1)\over 2T}\over 2E({\bf p}_1)} \Delta({\bf p}_1)\ d\tau_1  \ .
\label{bcs0} \eeq

   The analysis of solutions of the above nonlinear equations
 is greatly facilitated,
 if the interaction (\ref{longa}), taken at $\omega=0$, is
approximated by a $\delta$-function ${3\over 2}z^2\Gamma_a({\bf
q})\to f_a\delta({\bf q}-{\bf Q})$ \cite{schrief}, appropriate in
a domain of the momentum space, where the functions  $n({\bf p})$
and $\Delta_a({\bf p})$ change slower, than the amplitude
$\Gamma_a({\bf p}-{\bf p}_1)$. As a result, integrations cancel,
and we are left with
 \beq \xi({\bf p})=\xi^0({\bf p })+f_a  n({\bf p}-{\bf
Q},T), \qquad \xi({\bf p}-{\bf Q}) =\xi^0({\bf p}-{\bf Q})+ f_a
n({\bf p},T)\  , \label{eq22}
 \eeq
\beq \Delta({\bf p})= -f_a \Delta({\bf p}-{\bf Q}){\tanh(E({\bf
p}-{\bf Q})/ 2T) \over E({\bf p}-{\bf Q})}\ , \quad
\Delta({\bf p}-{\bf Q})=-f_a\Delta({\bf p}){\tanh(E({\bf p})/
2T) \over E({\bf p})}\   , \label{two} \eeq where the constant
$f_a>0$, is small compared to
 the band width $\omega_0$.  In  obtaining these equations we neglected the
term
$\Delta_r({\bf p})$.
 Setting here $T=0$, we arrive at a set of equations derived
 in Ref. \cite{zkc} in a different way.

If pairing correlations  are somehow suppressed,
Eqs.(\ref{two})
 are knocked out. Upon solving the two remaining Eqs.(\ref{eq22}) we find
 that
 in  case the van Hove points $(\pm\pi,0)$ and $(0,\pm\pi)$
are situated quite close to the Fermi line (FL), a portion of the sp
spectrum, adjacent to the van Hove points (vHP), turns out to be
flat \cite{zkc}. We shall see later that the flattening holds, if pairing
correlations come into play.

As seen from Eq.(\ref{two}), the gap $\Delta({\bf p})$ changes its
sign  going over to a neighbor vHP, as in the conventional D-pairing model, in
which the gap $\Delta_D({\bf p})\sim (\cos p_x-\cos p_y)\neq 0$
anywhere in the momentum space but the zone diagonals. However, in
contrast to this model,  nontrivial solutions of Eqs.(\ref{two})
exist only  in a domain $C$, boundaries of which are found by
combining two Eqs.(\ref{two}), which yields \beq f^2_a{\tanh (E({\bf
p})/2T)\over E({\bf p})}{\tanh (E({\bf p}-{\bf Q})/ 2T)\over
E({\bf p}-{\bf Q}) }=1 \ , \qquad {\bf p}\in C \ . \label{tem}
\eeq
 Otherwise  $\Delta_a\equiv 0$, and $E({\bf p})=|\xi({\bf p})|$,  as in the
Nozieres model \cite{noz,vol,schuck} with the effective long-range interaction
$\Gamma({\bf q})\sim \delta({\bf q})$.

In  overdoped compounds, the domain $C$ is made up of two  quite narrow
 stripes. The first, denoted further $C_F$ and described  by equation $\xi^0({\bf p})=0$, is adjacent
to the FL. The second, associated with the conjugate line (CL), is
determined  by equation $\xi^0({\bf p}-{\bf Q})=0$. In these
compounds,
 the FL and CL are well separated, and  when dealing with
 ${\bf p}\in C_F$ the energy
$E({\bf p}-{\bf Q})$  can be replaced by
$ |\xi^0({\bf p}-{\bf Q})|$, so that Eq.(\ref{tem}) is recast to
  \beq E({\bf p})=g({\bf p})\tanh(E({\bf p})/2T)  ,
\qquad {\bf p}\in C_F \  , \label{eqnoz} \eeq
with the coupling constant
$g({\bf p})=f^2_a/|\xi^0({\bf p}-{\bf Q})|$.

As $x$ drops, the FL and the CL   approach to meet each other at a
critical doping $x_m$. In  most of high-$T_c$ compounds, such as
Bi2212, Bi2201 etc., the FL is concave, while the CL,
respectively, convex, and the first meeting between these lines
occurs at the vHPs. Close to the vHPs,  boundaries of the $C$
domain are calculated combining Eq.(\ref{tem})
 with  relations $E({\bf p})=|\xi^0({\bf p})|$ and
 $E({\bf p})=|\xi^0({\bf p})+f_a|$, respectively, which yields restrictions
$-2f<\xi^0({\bf p})<f$. In this case, Eq.(\ref{tem}) is easily
solved, and close to the vHPs, the sp spectrum  turns out to be
 quite flat:   $E({\bf p},T=0) \simeq f_a$. We see that this value
 is significantly in excess of those obtained, if the FL and the CL
have no points of intersection. When
  the gap landscape  is drawn in the doping region
$x\sim x_m$, it comprises four "twin towers``, each one being  associated with
its own vHP. Each tower, whose height $\Delta_{vHP}(T=0,x_m)$, according
to Eq.(\ref{tem}),
   equals $f_a$, is  connected with its neighbors by narrow
"walls``. According to Eq.(\ref{eqnoz}), their  height
 drops towards  the zone diagonals, where the gap $\Delta$
vanishes. Thus, we infer that  the gap function  $\Delta(p_x,p_y)$
 attains its maximum $\Delta_{max}(T=0)=f_a$ at the vHPs.
 This  picture,  confirmed by
numerical calculations of
 Ref. \cite{zkc}, is in agreement  with the available experimental data \cite{camp}.

As T rises,   the region $C$, where $\Delta_a({\bf p})\neq 0$
 shrinks, the effect, found first in Ref. \cite{schuck} under investigation
of the Nozieres model \cite{noz}. Indeed, for points, fairly far away from
 the  vHPs, Eq.(\ref{eqnoz}) can be employed. Its
nontrivial solutions  exist only if $g({\bf p})>2T$. Since the
function $|\xi^0({\bf p}-{\bf Q})|$, identifying the energy
splitting between the FL and the CL,  rises, while the magnitude
of the function $g({\bf p})$ drops, respectively, as the vector
${\bf p}$ moves along the FL towards the zone diagonal,
 the shrinkage begins in the diagonal region at
 $ T_i(x)\simeq f^2_a/(2|\xi_{max}(x)|$, where
$\xi_{max}(x)$ is the bare sp energy, corresponding to
 the point of intersection between the CL and  the zone diagonal.
 With  further  $T$ increase, the shrinkage region is
 augmented, approaching the vHPs, where the gap $\Delta$ has its
 maximum value.
  Eventually,  the whole $C$
 domain shrinks into several symmetric points at the FL, closest
to  the vHPs. Recently, such a behavior of the gap landscape, the
so-called arc phenomenon \cite{camp} was
experimentally observed.
The  final shrinkage
temperature
 $T^*(x\sim x_m)$ is easily evaluated
 from Eq.(\ref{tem}). It is
$ T^*(x\sim x_m)\simeq f_a/2$, so that  the gap
 $\Delta_{max}(T=0)$ and $T^*$ are  connected with each other by
\beq \Delta_{max}(T=0)\simeq 2T^* \  , \qquad x\sim x_m\  ,\eeq
being in accord with the available experimental data \cite{shen,camp}.

It is worth noting that inside the C region, the behavior of
 $\Delta$ remains the same as  that in the BCS theory, since by retaining
 in Eq.(\ref{eqnoz}) the  leading terms, one obtains \beq
\Delta^2({\bf p}; T\to T^*) \simeq 12T^*(T^*-T), \qquad {\bf p}\in
C \ . \label{et} \eeq

Let us now turn to a rather rare case of the convex FL. Here
 the first intersection between the
FL and the CL occurs at the zone diagonals, and if one rotates
all the zone picture by the angle $\phi=\pi/4$,
 these points will coincide with the intersection points between the FL and the CL
 in the case of the concave FL.
 The analysis shows that this feature seems to hold
in dealing with all the solutions, including the gap landscape.
We shall revisit  this prediction of
 our model in a future paper.

So far we have neglected all the electron-electron interactions but the
longe-range one given by Eq.(\ref{longa}).
By involving an electron-phonon exchange, the most pronounced
out of the remaining ones, we
trigger, on one hand,
 a regular component
 $\Delta_r({\bf p})\neq 0$ anywhere in the momentum space. As a result,  the
Landau criterion for superconductivity, violated in the above
model at $T>T_i$, is now satisfied. Presumably, the magnitude of
$\Delta_r({\bf p})$ slowly varies with $x$, allowing us to
estimate it from highly
 overdoped compounds. Since in this case,
 $T_c$ is small, we infer that the impact  of $\Delta_r$
 on properties of
  the superfluid state  is insignificant.
On the other hand,  the electron-phonon exchange, specified by
the phonon  propagator $D(\omega,k)=k^2/(\omega^2-c^2k^2)$, gives
rise to
 a renormalization of the constant $f_a$, as well. The respective
 contribution  $\delta\Delta({\bf p})$ to  the gap $\Delta$ is given by the integral
 \beq \delta\Delta({\bf p})\sim \int D({\bf p}-{\bf p}_1,\omega_1)
  {\Delta_a({\bf p}_1)\over \omega^2_1-E^2({\bf p}_1)}\ d\tau_1{d\omega_1\over 2\pi i}  \ .
\label{def}\eeq
  Employing in the integral (\ref{def}) the "tower`` structure of the function
$\Delta_a({\bf p},x\sim x_m)$,
  one can decompose  ovewhelming   contributions into
  two: one from the same "tower`` and the other, from the neighboring one. The first
  contribution, proportional to
  the "tower`` range,  is small. When calculating the second one,
  where the momentum transfer ${\bf p}-{\bf p}_1$ is comparable to
  $p_F$,
   the propagator $D$ can be replaced by -1, yielding a
number, which suppresses the initial $f_a$ value. This interference may
be significant.

 Now we proceed to evaluation of
the superfluid density $\rho_s(T)$, expressed in terms of a
correlator of the velocities $\partial\xi^0({\bf p})/\partial {\bf
p}$.  Evaluation of this correlator in crystals with the help of
the Larkin-Migdal method \cite{lm,lar}
 yields:
 \beq \rho_s(T)=-{1\over 2}\int
{\partial\xi^0({\bf p})\over \partial p_i}\left[
{\partial n({\bf p},T)\over \partial  p_i}-
 {\partial f(E)\over
\partial E}{\partial\xi({\bf p})\over \partial p_i}\right] d\tau
\label{f1} \ , \eeq
 the function  $n({\bf p},T)$ being given by Eq.(\ref{np}).
  In ideal homogeneous Fermi gas, where $\xi(p)=p^2/2M-\mu$,
   Eq.(\ref{f1}) is converted into the
ordinary textbook formula.    Obviously,
  no  contributions to $\rho_s$ are made from  regions in
 momentum space, where the ratio $\Delta({\bf p})/T$ is negligible. Indeed,
 if $\Delta({\bf p})=0$,  the
 distribution $n({\bf p},T)$  is converted to
$n_F({\bf p},T)=(1+\exp(\xi({\bf p})/T))^{-1}$,
while the product
  $(\pr f(E)/\pr E)(\pr\xi({\bf p})/
 \pr p_i)\to
\pr n_F({\bf p},T)/\pr p_i$, and both the terms in Eq. (\ref{f1})  cancel
each other. As a result,
 at  $T>\Delta_r$,  contributions from regions, other than  the $C$
 domain,
 may be neglected.

A full examination of the formula (\ref{f1})  will be done elsewhere.
 Here we restrict ourselves to
the region of the  dopings $x\simeq x_m$ and temperatures $T\to
T^*\simeq f_a/2$. In this case, the ratio   $E({\bf p}\in C)/T$ is
small, and expansion of the terms in the integrand of
Eq.(\ref{f1})  yields
 $\pr n({\bf p},T)/\pr p_i\simeq
 -(\pr \xi({\bf p})/\pr p_i)/( 4T) +\xi^2({\bf p})(\pr \xi({\bf p})/
 \pr p_i)/( 16T^3)+
 \Delta^2({\bf p})(\pr \xi({\bf p})/\pr p_i)/(48T^3)$ and
 $\pr f(E)/ \pr E\simeq -1/(4T)+E^2({\bf p})/( 16T^3)$. After
 cancelling  similar terms and employing relation (\ref{et}), we are  left with
 \beq
\rho_s(T\to T^*)\simeq{1\over 48} \int_C{\pr\xi^0_{{\bf
p}}\over\pr p_i}\Delta^2({\bf p}){\pr\xi({\bf p}) \over \pr p_i}
d\tau\simeq
 \alpha n(T^*-T)^2/(T^*)^2\  , \label{supd}
\eeq
 where the numerical factor  $\alpha$ is of the order of $ 10^{-2}$.
As we shall see, such a suppression of $\rho_s(T\to T^*)$,
 results in a marked distinction between  the
 critical temperature  $T_c$ for termination of  superconductivity and the
 temperature $T^*$ for vanishing  of the gap  $\Delta$.
The  reason for that is
 a great  diversity in the gap  values, which, as we have seen,
  results in   the temperature  shrinkage  of  the domain of
 integration  over  momentum space in Eq.(\ref{f1}).

Strictly speaking, in two-dimensional systems, the temperatures
$T^*$ and $T_c$
 never coincide because of the Berezinskii-Kosterlitz-Thouless (BKT) phase
transition \cite{kt,al,ek},  terminating superconductivity due to  a spontaneous
generation of an infinite number of
vortices. This transition
always occurs before vanishing of   the gap $\Delta$.
The BKT temperature, being, in fact, the critical temperature $T_c$,
 is given by the equation \cite{kt}
  \beq \pi T_c= 2 \rho_s(T_c)\ .\label{bktc} \eeq
 In conventional superconductors, where  $\rho_s(T\to T^*)\sim
 n(T^*-T)/T_c$, and, hence,
   $ (T^*-T_c)\sim {T_c^2/
\epsilon^0_F}$, the ratio $T_c/\epsilon^0_F$ does not exceed 0.2\%.
 However, in two-dimensional electron compounds
with the doping $x\sim x_m$, the situation is different. Indeed,
upon inserting the result (\ref{supd}) into Eq.(\ref{bktc}),  one
obtains
\beq T_c=\alpha_m\epsilon^0_F(T^*-T_c)^/(T^*)^2  \ , \label{bkt1} \eeq
where the  factor  $\alpha_m\simeq 10^{-2}$.
 In high-$T_c$ superconductors, the ratio $T_c/\epsilon^0_F$
is of order of $10^{-2}$, and, hence,
the ratio  $\tau=(T^*-T_c)/T^*$ may attain values comparable to unity.

In conclusion, we have evaluated the effect of critical antiferromagnetic
fluctuations on electron spectra and superfluid densities of
superfluid states of overdoped and
optimally doped high-$T_c$ compounds. In
 underdoped electron systems, the  situation is more complicated due to the
emergence  of a branch of low-lying collective excitations, whose
contribution to properties is significant \cite{norm}.
The interplay between these oscillations and critical fluctuations
 in underdoped compounds will be studied  a separate paper.

This research was supported in part by the National Science Foundation
under Grant No.~PHY-0140316, by the McDonnell Center for the Space
Sciences, and by Grant No.~00-15-96590 from the Russian Foundation
for Basic Research (VAK).
  We thank L. P. Gor'kov, G. Kotliar, A. I. Lichtenstein,
M.~R. Norman, N.~E.~ Zein and M.~V.~ Zverev
  for valuable discussions.  We also thank  ITF
( Santa Barbara, USA) for the kind hospitality.

\end{document}